\documentclass[a4paper,fleqn,usenatbib]{mnras}

\usepackage[T1]{fontenc}
\usepackage{ae,aecompl}

\usepackage{natbib}
\usepackage{graphicx}
\usepackage{enumitem}
\usepackage{calc}
\usepackage{lipsum,adjustbox}
\usepackage{subcaption}
\usepackage{tikz}
\usepackage{amsmath}	% Advanced maths commands
\usepackage{amssymb}
\usepackage{wasysym}
\usepackage{txfonts}
\usepackage{array}
\usepackage{threeparttable}
\usepackage{float}
\title[Asteroseismic Classification with ConvNets]{Deep Learning Classification in Asteroseismology}

\author[Hon et al.]{
	Marc Hon,$^{1}$\thanks{E-mail: mtyh555@uowmail.edu.au}
	Dennis Stello,$^{1,2,3}$ 
	and Jie Yu$^{2}$
	\\
	$^{1}$School of Physics, The University of New South Wales, Sydney NSW 2052, Australia\\
	$^{2}$Sydney Institute for Astronomy (SIfA), School of Physics, University of Sydney, NSW 2006, Australia\\
	$^{3}$Stellar Astrophysics Centre, Department of Physics and Astronomy, Aarhus University, Ny Munkegade 120, DK-8000 Aarhus C, Denmark\\
}

\date{Accepted XXX. Received YYY; in original form ZZZ}

\pubyear{2017}

\begin{document}
	\label{firstpage}
	\pagerange{\pageref{firstpage}--\pageref{lastpage}}
	\maketitle
	
	\begin{abstract}
		In the power spectra of oscillating red giants, there are visually distinct features defining stars ascending the red giant branch from those that have commenced helium core burning. We train a one-dimensional convolutional neural network by supervised learning to automatically learn these visual features from images of folded oscillation spectra. By training and testing on \textit{Kepler} red giants, we achieve an accuracy of up to 99\% in separating helium-burning red giants from those ascending the red giant branch. The convolutional neural network additionally shows capability in accurately predicting the evolutionary states of 5379 previously unclassified \textit{Kepler} red giants, by which we now have greatly increased the number of classified stars.

	\end{abstract}
	
	\begin{keywords}
		asteroseismology -- methods: data analysis -- techniques: image processing -- stars: oscillations -- stars: statistics
	\end{keywords}
	
	%%%%%%%%%%%%%%%%%%%%%%%%%%%%%%%%%%%%%%%%%%%%%%%%%%
	
	%%%%%%%%%%%%%%%%% BODY OF PAPER %%%%%%%%%%%%%%%%%%
	
	\section{Introduction}
	
	A key concept in determining stellar ages of red giants is distinguishing the evolutionary state i.e. classifying between hydrogen-shell burning stars ascending the red giant branch (RGB) and those that have commenced core helium burning (HeB). Space missions such as \textit{Kepler} \citep{Borucki2010} have provided an enormous quantity of red giant oscillation spectra. The upcoming TESS mission \citep{TESS} will further increase the amount of data, making manual or semi-automatic classification of the population class of each star infeasible. Automated methods do exist, however considerable effort is required for defining and acquiring features such as the observed period spacing $\Delta P$ \citep{Bedding_2011, Stello2013}, the asymptotic period spacing $\Delta\Pi_1$ \citep{Bedding_2011, Mosser_2014, Vrard2016}, or the structure of mixed modes \citep{Elsworth2016} in order to separate the populations. Furthermore, these methods require relatively high signal-to-noise data.
	
	Here, we present a deep learning method that allows spectral features to be learnt by the machine using convolutional neural networks. These are machine learning methods that mimic biological neuron structures, aimed towards feature detection in data \citep{Fukushima_1980}. They have achieved significant success over the past few years in computer vision methods such as image recognition \citep{Kriz2012}, and even facial recognition \citep{Garcia_2004}.
	
	We introduce the concept of representing the oscillation frequency spectra of stars as `images'. While spectra are represented as a 1D array of values and are not images in the conventional 2D sense, we interpret the spectra in an image-like fashion in such a way that we wish to identify the spatial structure of the spectra in frequency as visual features; as seen by `an expert eye'. Hence, these images are simple representations of the power excess, which contain visual features for RGB-HeB classification that the neural network can `see' and hence learn from. By learning from an existing set of classified red giants based on $\Delta\Pi_1$ measurements, we use 1-D convolutional neural networks as a form of supervised machine learning aimed to automatically learn features separating RGB from HeB stars in order to  make fast yet accurate classification predictions on vast amounts of unclassified stars.

	\section{Methods}
	
	Here we describe the preparation of the image representation known as a \textit{folded spectrum}, along with an overview of convolutional neural networks, and the construction of a deep learning classifier to classify the image representation.
	\begin{figure}
		\centering
		\includegraphics[width=\linewidth]{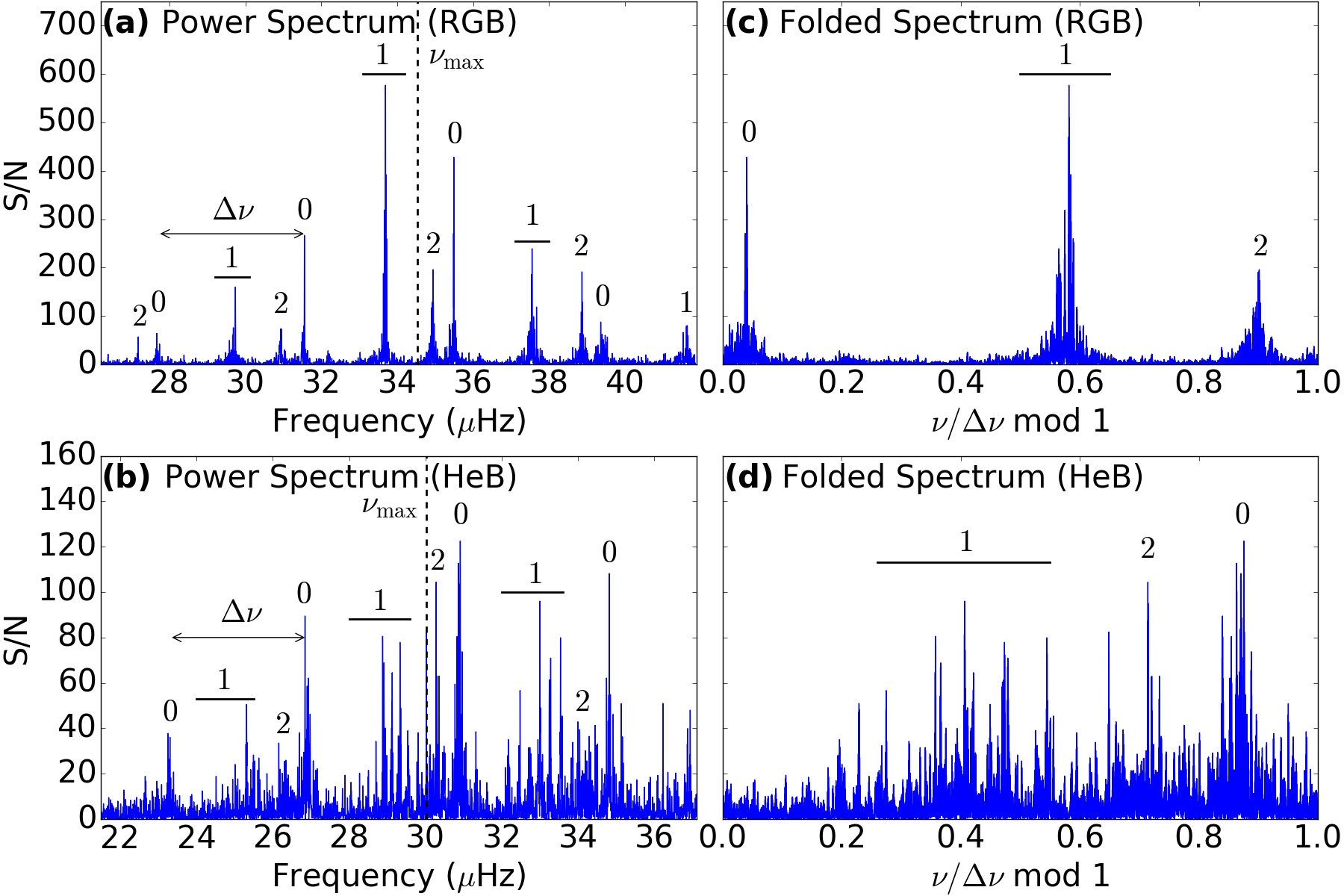}
		\caption{Comparisons between an RGB star KIC 11293804 (top), with a HeB star KIC 5810333 (bottom), both having large frequency spacing $\Delta \nu \simeq 3.92 \mu$Hz. (a) and (b) are the original power spectra, while (c) and (d) are \textit{folded spectrum} image representations. The oscillation modes are labelled by their degree $l$, while the frequency of maximum oscillation power, $\nu_{\mathrm{max}}$, is indicated by the dashed vertical line.}
		\label{popcomp}
	\end{figure}
	
	%4755
	\subsection{Data}
	\label{sec:data}
	We obtain the evolutionary state classifications of 5673 \textit{Kepler} stars based on automated asymptotic period spacing measurements by \citet[][hereafter Vrard]{Vrard2016}, and add 335 stars from the classification by \citet[][hereafter Mosser]{Mosser_2014} that are not already in Vrard's sample, to a total of 6008 stars. We then assign RGB stars with the binary class 0 and HeB stars with class 1. About 30\% of stars in the dataset are RGB stars. We randomly choose 1008 stars as test data, with the remaining 5000 stars for training. Additionally, we have an \textit{unclassified set} comprising 8794 \textit{Kepler} red giants that are known to oscillate but have not been given classifications by Vrard or Mosser. We want to predict the population labels of all stars in our unclassified set using our trained neural network.
	
	\subsection{Image Representation}
	\label{DataRep}
	\begin{figure}
		\centering
		\includegraphics[width=\linewidth]{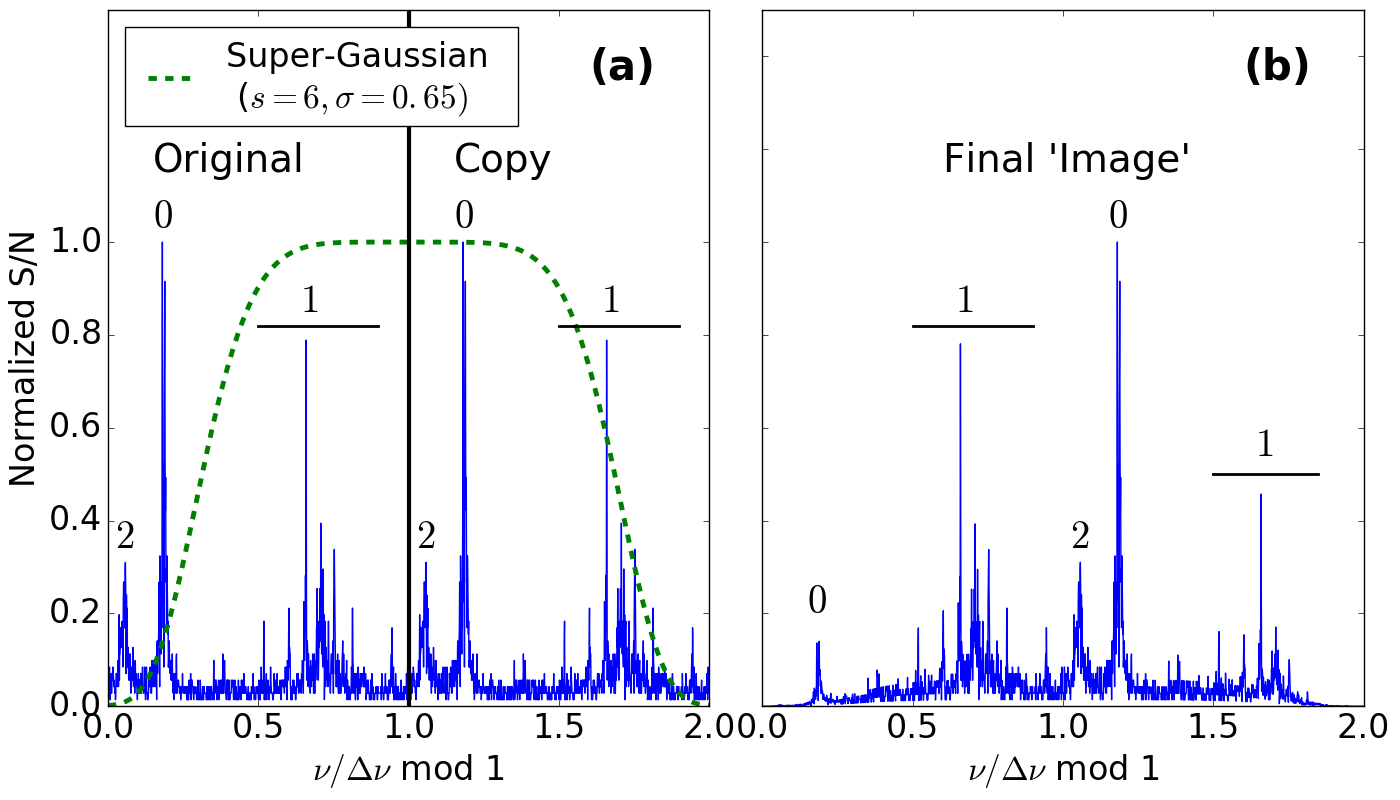}
		\caption{(a) Appended and normalized folded spectrum of KIC 10790301 with mode identification. $s$ is the shape parameter or order (=2 for standard Gaussian), while $\sigma$ is the standard deviation of the super-Gaussian weight. The solid vertical line separates the original image from its appended copy. (b)  Resulting image from application of the super-Gaussian weight to the spectrum in (a).}
		\label{supergauss}
	\end{figure}
	As our image representation, we define the \textit{folded spectrum} as the $4\Delta \nu$-wide power spectrum segment centred at $\nu_{\mathrm{max}}$, folded by a length of $\Delta \nu$ (see Figures \ref{popcomp}a, c). The spectra and values for $\Delta \nu$ and $\nu_{\mathrm{max}}$ were derived from end-of-mission \textit{Kepler} data using the SYD pipeline \citep[][Yu et al., in prep.]{SYD}. Because the neural network requires a fixed input array length, we bin each folded spectrum into 1000 bins.
	
	A comparison of spectral image representations between RGB and HeB stars are shown in Figures \ref{popcomp}c, d. RGB stars clearly exhibit acoustic modes that are highly localized (Figures \ref{popcomp}a, c) while HeB stars show broader mode distributions particularly for non-radial modes because of the stronger coupling between core and envelope (Figures \ref{popcomp}b,d) \citep{Dupret2009, Grosjean_2014}. With acoustic resonances less localized, HeB spectral representations notably have greater visual complexity as compared to RGB spectra. Besides the structure of modes, the location of the $l=0$ mode, represented by $\epsilon$, can be a strong indicator in distinguishing population classes \citep{Kallinger2012}. However, $\epsilon$ is not the sole feature that is used to recognize population classes from an image. The lack of a clear boundary separating the two evolutionary states shown by the observed spread in $\epsilon$ \citep{Kallinger2012} and from theoretical studies \citep{CD2014} makes $\epsilon$ unsuitable as a sole selection criterion. However, information about $\epsilon$ complements features extracted from mixed modes in the image.
	
	As image pre-processing, we normalize each spectrum by its max power value. Then, to avoid edge effects, we append the image with a copy of itself and apply a super-Gaussian (higher-order Gaussian) weight function as shown in Figure \ref{supergauss}.
	\begin{figure*} %Reference for ConvNet section
		\centering
		\subcaptionbox{An elementary input-output connection in a neural network layer. Each neuron (circle) holds a real number. The activation function, $f$, maps the dot product of the weight vector, \boldmath$\mathrm{w}$, and the input vector, \boldmath$\mathrm{x}$, into a non-linear output.}[.6\columnwidth]{\label{Nodepic}
			\includegraphics[width=.65\columnwidth]{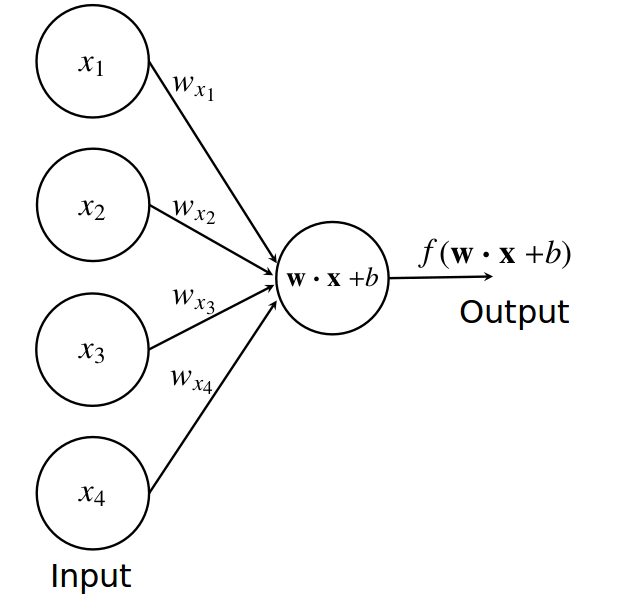}
		}\hfill
		\subcaptionbox{Three fully-connected feedforward neural network layers with one hidden layer and two output neurons.}[.4\columnwidth]{
			\includegraphics[width=.45\columnwidth]{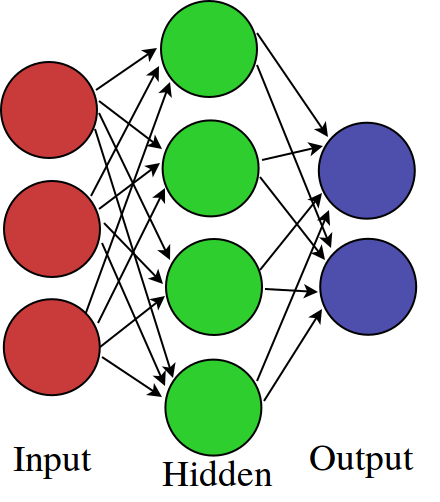}
			\label{Densepic}}\hfill
		\subcaptionbox{Convolutional filtering applied to a one-dimensional input. A kernel with weights ($w_1, w_2, w_3$) is moved over input neurons, three at a time. %Each kernel corresponds to a feature map, with a fixed set of weights 'shared' across the input.%
			The weights across inputs here are fixed as a generic set of values for the feature map, whereas in (a), each weight may be unique.}[1.\columnwidth]{
			\includegraphics[width=1.05\columnwidth]{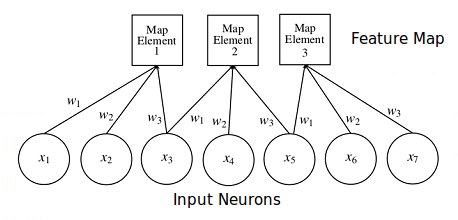}
			\label{Convpic}}\hfill
		\caption{A general schematic of neural network layers.}
		\label{NNpic}
	\end{figure*}

	\subsection{Convolutional Neural Networks}
	
	An artificial neural network is a mathematical representation of a biological neuron network (Figure \ref{NNpic}a). Mathematical neurons contain real numbers and connect to other neurons in subsequent levels (network layers) by mathematical operations in order to form a network capable of computing solutions to complex problems. The total input from one layer to a neuron in the next network layer is given by \boldmath$\mathrm{w \cdot x}$, where $\mathrm{x}$\unboldmath$=(x_{0}, x_{1}, x_{2}, x_{3},..., x_{n})$ is an input vector with $n$ number of features from the input layer (represented by the number of neurons in the input layer in Figure \ref{NNpic}a), with $x_0 = 1$. In our study, \boldmath$\mathrm{x}$ is the power in each frequency bin of the image at the very first input layer. In subsequent layers, \boldmath$\mathrm{x}$ will become manipulated representations of the original input image.  The weight vector, \boldmath$\mathrm{w}$\unboldmath$=(w_{0}, w_{1}, w_{2},...,w_{n})$, links each input to a neuron in the subsequent layer, with $w_{0}$ known as the input bias, $b$, which is analagous to the intercept in a linear regression.
	
	The total input is linear, however it is passed through a non-linear \textit{activation function} \citep{Blatt62}, $f$, such that the network becomes capable of approximating complex non-linear representations. In this study, we use the \textit{rectified linear unit} activation function $f(x)=\mathrm{max}(0,x)$ for every neural network layer except the output layer. This function is suited for feature learning in neural networks \citep{Nair2010}.  A common design in neural networks is to stack multiple layers and have the input pass through consecutive intermediate or \textit{hidden} layers to reach the output layer (Figure \ref{NNpic}b). Such a design is known as a \textit{feedforward} neural network, because inputs are computed and fed forward through the network to the output layer. A simple feedforward neural network has fully-connected layers, such that each neuron in a layer is fully connected to the neurons in the following layer. Each neuron connection for a fully-connected layer is permitted to have distinct weights.
	
	Convolutional neural networks \citep{LeCun1998} are a variant of feedforward neural networks in which the layer connections are constrained. This constraint comes in the form of \textit{weight sharing}, where weights across neurons within a layer are constrained to only a fixed set of values, known as a \textit{filter}. The content of the neurons in a convolutional layer is computed by sliding this filter across neurons in that layer (Figure \ref{NNpic}c). Hence, the filter is analagous to a kernel convolution. By using a fixed-length filter as weights instead of allowing each neuron connection to have their own distinct weights, features across a local `patch' of data, for instance, the shape of an $l=1$ mixed mode in Figure \ref{supergauss}b, can be learned to be detected by the network. Detected features across a layer are computed and then stored in a \textit{feature map} \citep{Rumelhart1988a}. Feature maps act as the hidden layer for convolutional layers, with the exception that whereas fully-connected layers usually have a single array for its hidden layer, convolutional layers typically have more than one feature map because the detection of multiple features require multiple feature maps. A feature map is given by:
	\begin{equation}
	g^{(l)} = f \bigg(\sum_{i=0}^{m}\boldmath \mathrm{w}^{(l)} \circ \ \mathrm{x}_{i}\bigg),
	\label{convEq}
	\end{equation}
	with $m$ denoting the number of stars in the data set and $l$ being the feature map index. Each feature map can be said to represent a detected local spatial feature of the data. However, in image recognition, as a greater number of convolutional layers are added, the outputs of deeper layers often become increasingly difficult to interpret from a human visual perspective \citep{Zeiler_2014}.
	
\begin{figure*}
		\centering
		\includegraphics[width=0.8\linewidth]{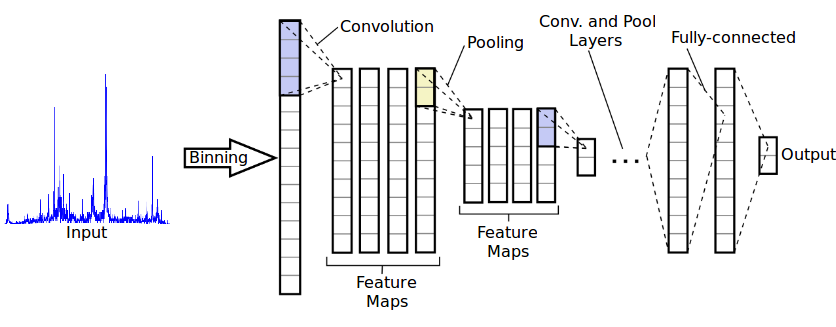}
		\caption{General schematic of a convolutional neural network. The length of each array in this general schematic is arbitrary.}
		\label{ConvSchematic}
	\end{figure*}

	A pooling layer is commonly applied after convolutional layers. Pooling reduces the length of convolutional layer outputs by clustering adjacent neurons into a fewer number of neurons over a specific function that compares the neuron values. To illustrate this, in our study, we use pooling layers that apply a 4-neuron \textit{max-pooling}, which reduces a collection of 4 adjacent neurons into 1 neuron by selecting the neuron with the maximum value among them. In principle, this achieves a form of local spatial invariance within layers \citep{Bengio2013}, while reducing the complexity of the neural network at the same time because fewer weights are required for a smaller layer size. Following feature detection by multiple iterations of convolution and pooling layers, it is common for a fully-connected layer to be connected for a final layer of computation before the output layer. A general structure of a convolutional neural network is illustrated in Figure \ref{ConvSchematic}. 
	
	\subsection{Learning and Optimization}
	\label{Learning}
	The objective of the convolutional neural network is to learn a particular set of weights from a training set that minimizes the error in approximating a ground truth ${y}$ with a predicted output $\hat{y}$. In binary classification, $y$ are binary values (associated with the binary classes; 0 for RGB and 1 for HeB) while $\hat{y}$ are 2-element vectors with output scores for each class. We use the normalized exponential function known as the \textit{softmax} function at the output layer, such that each output neuron contains the value:
	\begin{equation}
	p(y=j|\mathrm{\textbf{x}}) =\frac{e^{\mathrm{\textbf{x}}\cdot\mathrm{\textbf{w}}_j}}{\sum_{k=1}^{K}e^{\mathrm{\textbf{x}}\cdot\mathrm{\textbf{w}}_k}},
	\label{softmax}
	\end{equation}
	which defines the score of class $j$ out of $K=2$ classes. These scores are similar to probabilities, however, probability calibration is usually required to express these scores in terms of actual population probabilities. Since such work is beyond the scope of this study, it is sufficient to interpret the output scores as the prediction \textit{likelihood} of a certain population class for a star. A score close to 1 indicates a high predicted likelihood for a HeB star, while a score close to 0 implies a high likelihood for an RGB star. We use a score threshold of 0.5 to assign the dominant population label of a target, such that a predicted score close to 0.5 implies a lack of classifier confidence in identifying the star's population label as both populations 'appear' to be equally likely to the classifier. This score threshold has been found to provide a good separation of populations in our classifier.
	
	A suitable error function to minimize is the \textit{cross-entropy} or log-loss \citep{Murphy2012}, related to the difference between a true value, $y$, with a predicted value, $\hat{y}$. Thus, the log-loss is the measure of similarity between $y$ and $\hat{y}$. We compute the error, $E$,  of the network by taking the average of all cross-entropies of the network's predictions on $m$ stars, such that
	\begin{equation}
	E(\mathbf{y, \hat{y}}) = -\frac{1}{m} \sum_{i=1}^{m} \bigg[y_i\log\hat{y}_i + (1-y_i)\log(1-\hat{y}_i)\bigg].
	\end{equation}
	
	The error minimization uses a gradient descent algorithm, where the error derivatives with respect to a layer's weights are calculated from the end output and backpropagated to previous layers \citep{Rumelhart1986}. Layer weights are then updated sequentially over small subsets (mini-batches) of the training examples, which speeds up learning compared to updating over the full training set at once because it approximates the gradient and the error surface curvature \citep{LeCun1998a}. We use a mini-batch size of 128 to train our network.  During the training process, multiple passes of feed-forward and backpropagation are iterated until the error converges to a minimum. By training a convolutional neural network with a suitable optimization objective, we obtain a \textit{classifier} (a fixed network with a trained set of fixed weights) that takes in our image as input and outputs a two-element vector $\hat{y}$ with the elements as the RGB and HeB likelihoods, respectively.
	
	\subsection{Classifier Structure and Hyperparameters}
	\begin{figure}
		\centering
		\includegraphics[width=.55\linewidth]{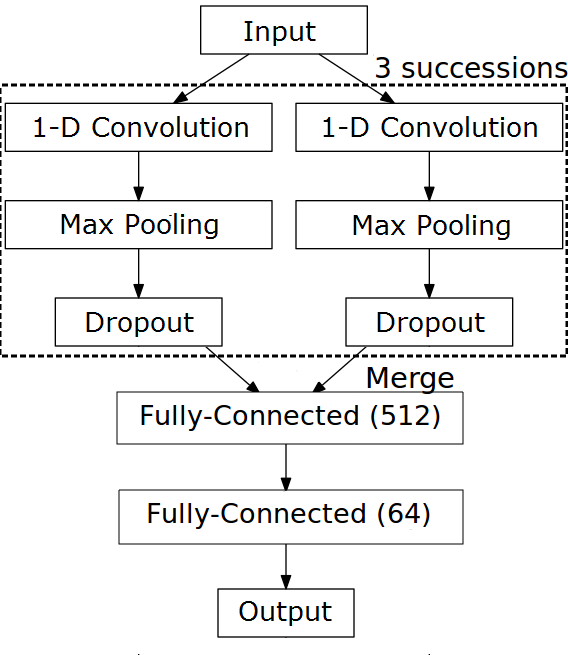}
		\caption{Structure of our convolutional neural network. The number in brackets for fully-connected layers is the input array length. The structure outlined by the dashed box is run three times in succession before merging the layer outputs by tensor concatenation.}
		\label{architecture}
	\end{figure}
	A deep learning classifier usually has multiple stacks of neural network layers on top of one another as its structure. This structure contains a vast combination of free parameters (\textit{hyperparameters}), which have to be empirically determined. Our classifier structure (Figure \ref{architecture}) uses two parallel yet identical convolutional layer stacks such that both stacks see the same input. The sequence of Convolution-Pooling-Dropout (see explanation of dropout below) is carried out 3 times in succession before merging into a fully-connected layer. We use a filter size of 32 in each convolutional layer with $2/4/8$ feature maps for iterations 1/2/3. This structure is defined by choosing the simplest, computationally cheap combination of structure and hyperparameters that has the best metric performance on a \textit{hold-out validation set}. This set is part of the initial training data that is now left out in training for classifier structure selection.
	
	To prevent overfitting on training data, we use \textit{dropout} layers \citep{Hinton2012}, which randomly sets layer outputs to zero with probability $p_{\mathrm{drop}}$. This prevents layer outputs from memorizing the training data. We performed dropouts with $p_{\mathrm{drop}}=0.6$ after each convolutional layer and constrain the norm of the layer's weight vector to a value of 2. This combination has been shown to be effective at preventing overfitting \citep{Sriva2014}. %To ensure proper error convergence, we initialize layer weights with a \textit{Glorot normal} distribution \citep{Glorot2010}, which is a Gaussian distribution normalized by the number of inputs to a layer node (\textit{fan-in}) added to the number of outputs of that same node (\textit{fan-out}).
	We constructed the classifiers with the Keras library \citep{Keras} built on top of Theano \citep{Theano}. With a Quadro K620 GPU, training with 200 training iterations requires approximately 15 minutes, while predicting on thousands of new stars takes only a few seconds.
	
	\section{Results}
	
	\subsection{Classifier Performance}
	We report metrics on the mean of 10 separate hold-out validation sets, with each set having one-tenth of the training set in size (10-fold cross-validation), and on the test set. The metrics used to describe classifier perfomance are defined as follows:
	
	\begin{description}[]
		\item[Accuracy:] The number of correct predictions out of all predictions. 
		\item[Precision(P):] For a class, the ratio of correct predictions to \textit{all made predictions} towards that same class. Here it is the classifier's ability to not label a HeB star as an RGB star.
		\item[Recall(R):] For a class, the ratio of correct predictions to \textit{all stars} truly in that same class. Here it is the classifier's ability to find all HeB stars.
		\item[F1 Score:] The harmonic mean of precision and recall, defined by $2\frac{P\times R}{P + R}$, with 1 as a perfect score.
		\item[ROC AUC:] \textbf{R}eceiver \textbf{O}perating \textbf{C}haracteristic's \textbf{A}rea \textbf{U}nder \textbf{C}urve, which measures the classifier's average performance across all possible score thresholds. Has a value of 1 for a perfect classifier \citep{Swets2000}. 
		\item[Log Loss:] Negative logarithm of prediction scores i.e. the \textit{cross entropy}. Measures how well prediction scores are calibrated with an ideal value of 0.
		
	\end{description}
	Precison, recall, and the F1 scores complement accuracy in cases like ours where the population ratio is far from 50:50, while ROC AUC evaluates the overall classifier performance. Log loss reports the performance of class score outputs, with confident predictions rewarded low error when correct while penalised heavily otherwise.
	\begin{table} 
		\centering
		\caption{Metrics over the mean of 10-fold cross-validation (CV) and over the test set.
		}
		\label{results}
		\begin{threeparttable}
			\begin{tabular}{|l|c|c|}
				\hline
				Dataset & CV ($\pm 1$ std.) & Test\\
				\hline
				Accuracy&0.982 $\pm$ 0.005 &0.990\\
				Precision&0.982 $\pm$ 0.005 &0.990\\
				Recall&0.982 $\pm$ 0.005 &0.991\\
				F1 Score &0.982 $\pm$ 0.005&0.991\\
				ROC AUC &0.998 $\pm0.002$ &0.996\\
				Log Loss &0.055 $\pm0.020$ &0.044\\
				\hline
				
			\end{tabular}
		\end{threeparttable}
	\end{table}
	\begin{figure}
		\centering
		\includegraphics[width=1.02\linewidth]{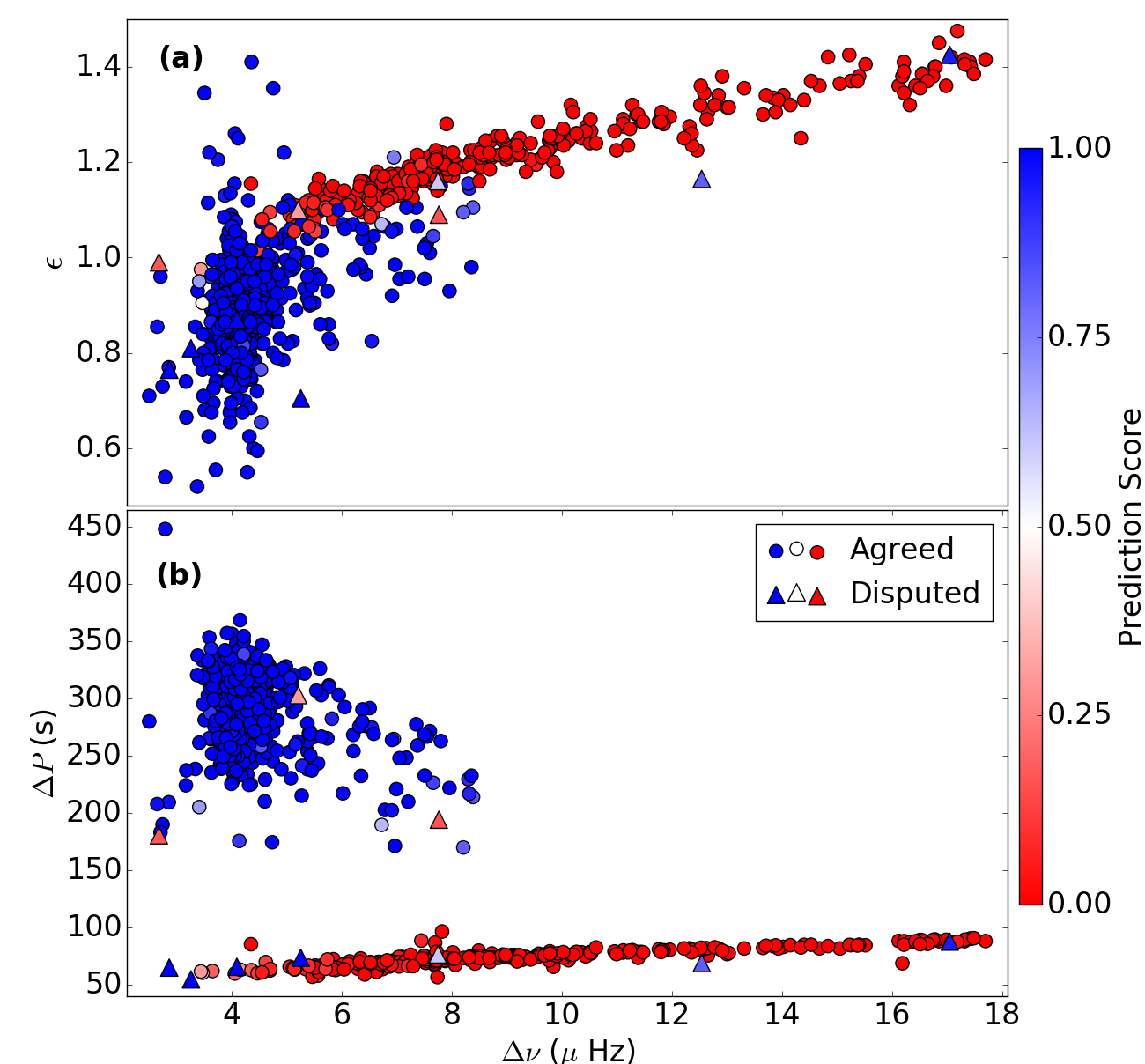}
		\caption{(a) $\epsilon - \Delta \nu$ diagram and (b) $\Delta P - \Delta \nu$ diagram of test set predictions. The colour corresponds to the score values of predictions, with deeper colors corresponding to a greater confidence towards a particular class (0 for RGB and 1 for HeB). Our classifications are available online <link>.}
		\label{dnudP}
	\end{figure}
	As observed in Table \ref{results}, the accuracy of the classifier on the cross-validation sets is generally above 98\%. On the test set where the classifier benefits from training on the entire training set, the classifier is capable of classifying with a 99\% accuracy and suffers a lower log loss.
	Having high values of precision, recall, and F1 score also indicates that the classifier is not heavily biased in predicting a particular population class that would not reflect the true population ratio.
	
	Figure \ref{dnudP} shows the test set results in $\epsilon - \Delta \nu$ \citep{Huber_2010,White2011} and $\Delta P - \Delta \nu$ \citep{Bedding_2011} diagrams. We derived the $\epsilon$ values using the method described in \citet{Stello_2016b,Stello_2016a}. One can see that `disputed' predictions, namely predictions that are not in agreement with the "truth" labels from Vrard or Mosser, are more concentrated towards the low-$\Delta \nu$ regions. The classifier appears to be confident in most of its predictions (deep red and deep blue symbols), while most of the uncertain predictions are disputed. Upon inspection of the spectra of the 10 disputed stars, we visually verify that four of them, all with $2.9\mu$Hz $<\Delta \nu$ $< 5.2 \mu$Hz, had incorrect ground truth labels. Another four are confirmed to be due to the classifier's inaccuracy. These stars have $\Delta \nu$ $> 7.0 \mu$Hz in the diagrams. The final two stars are "high" luminosity red giants with $\Delta \nu < 2.9 \mu$Hz. Visual inspection was inconclusive as the spectrum of one had suppressed dipole modes with a moderate level of noise, while the other appeared much like an RGB star but was previously given a late HeB classification as its ground truth. From theory, we do not expect to see a clear difference between RGB and late HeB stars because of the lack of coupling between core and envelope in such stars \citep[][their Fig. 4b]{Stello2013}. 
	
	\subsection{Classifying the Unclassified Set}
	\begin{figure}
		\centering
		\includegraphics[width=1.05\linewidth]{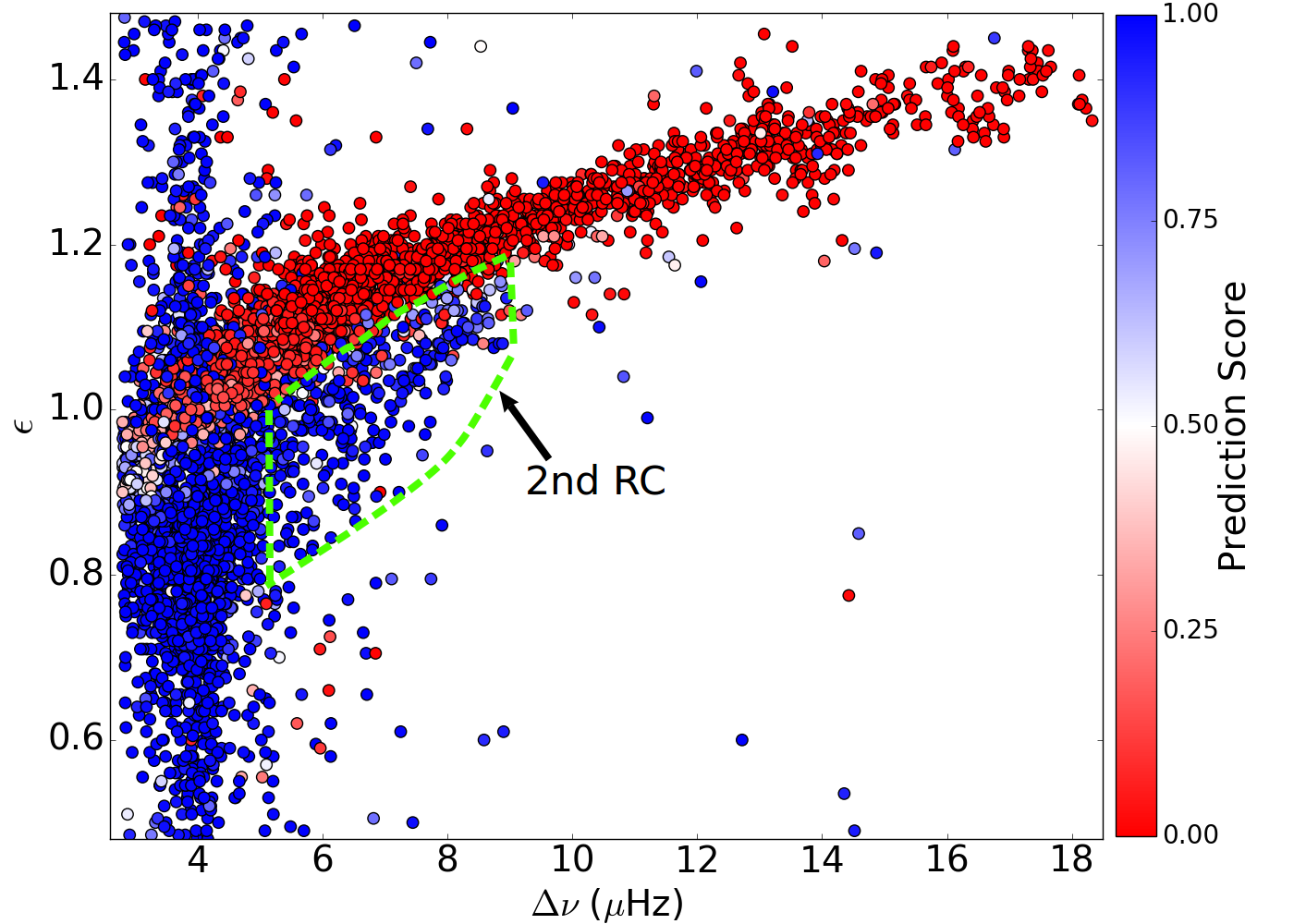}
		\caption{$\epsilon - \Delta \nu$ diagram of the unclassified set for stars with $\Delta \nu \apprge 2.8\mu$Hz. The colour scheme is similar to that of Figure \ref{dnudP}. The region of the secondary clump stars is also indicated. The classifications for this plot are available online <link>.}
		\label{dnuepsilon}
	\end{figure}
	We now use our trained classifier to predict the evolutionary state of the unclassified set (Figure \ref{dnuepsilon}). It can be seen that the predictions reflect the $\Delta \nu - \epsilon$ relation of RGB stars \citep{Kallinger2012} well for the entire $\Delta \nu$ range spanned by the training set ($2.8\mu$Hz $\apprle \Delta \nu \apprle 18\mu$Hz). The secondary clump of HeB stars is seen in the diagram with $\Delta \nu \simeq 6-9 \mu$Hz below the RGB $\Delta \nu - \epsilon$ relation. In addition, the predictions also clearly show the HeB population at $\Delta \nu \simeq 3-4\mu$Hz, with $\epsilon$ values mostly ranging about 0.7 to 1.0. In Figures \ref{dnudP}a and \ref{dnuepsilon}, stars with $\Delta \nu \simeq 4\mu$Hz and $\epsilon \apprge 1.2$ are low $\epsilon$ stars that have `wrapped around' vertically in the diagram. 
	
	The classifier has mostly very confident predictions, shown by the vast majority of predictions in Figure \ref{dnuepsilon} having deep shades of red and blue for RGB and HeB predictions, respectively. The exception to this is at the region of intersection between RGB and HeB populations at $2.8\mu$Hz $\leq\Delta\nu\apprle4.0\mu$Hz and 0.9 $\apprle\epsilon\apprle$1.0. The paleness of the symbol colours in this region indicate a prediction uncertainty, which we attribute to the scarcity of stars in our training set with $\Delta\nu$ and $\epsilon$ values within this range.
	
	Despite the predictions capturing the general distribution of red giant populations within the $\Delta \nu - \epsilon$ diagram , the classifier has its limitations from classifying based on image representation alone. For instance, it does not explicitly discriminate between frequency spacings, such that it can erronously predict HeB stars at high $\Delta \nu$ ($\Delta \nu \apprge 9\mu$Hz), where no HeB stars exist. However, only a very small fraction of predictions are subject to this inaccuracy. Another important limitation of these predictions is imposed by the parameter range of the training data. Our training data only includes RGB stars down to $\Delta \nu \simeq 2.8\mu$Hz, hence we infer that the reliability of the classifier predictions also holds to a similar $\Delta \nu$ threshold. Due to this, we do not provide classifications for 1139 red giants with $\Delta \nu < 2.8\mu$Hz in our unclassified set, leaving behind 7655 red giants with $\Delta \nu \geq 2.8\mu$Hz from the initial 8794 in the unclassified set. Out of the remaining 7655 red giants, we find that 517 stars have been previously classified by \citet{Stello2013}, from which 43 of their predictions do not agree with ours. We visually verify that they had incorrectly labelled approximately half of the disputed stars as HeB, with a majority of them having $\Delta \nu \simeq 5.5 \mu$Hz. In addition, 1991 stars with $\Delta\nu \apprge 2.8\mu$Hz have been classified by \citet{Elsworth2016}, from which 232 stars are overlapping with the stars classified by \citet{Stello2013}. Our classifier produces disputing predictions for 195 out of the 1991 stars classified by \citet{Elsworth2016}. These disputes are mostly for those stars that they have predicted as RGB in the range of $2.8\mu$Hz $\apprle \Delta \nu \apprle 5\mu$Hz, which we find are split in roughly equal numbers into those that our classifier predicts correctly, those that our classifier predicts incorrectly, and those where we are uncertain of the true population by visual inspection. In the end, we produce new classifications for 5379 stars not classified before by any other method.  In future work, we will develop methods to improve the generalization of classifier predictions across a greater range of asteroseismic parameters. In addition, we will also look into spectral features of high $\Delta \nu$ RGB stars, which can deceive the classifier to incorrectly predict the star as HeB.
	
	\section{Conclusions}
	
	We have developed a convolutional neural network to perform fast and efficient classification of red giant stars into those ascending the red giant branch and into those that have commenced core helium burning. We presented folded oscillation spectra as images, which contain visual features that are learned by the convolutional neural network. Training and testing on \textit{Kepler} data yielded a 98\% cross-validation accuracy and a 99\% test set accuracy, benchmarked against classifications based on asymptotic period spacing measurements. Out of the predictions that were in conflict with the `ground truth', most scenarios of classifier inaccuracy were limited to the intermediate to high $\Delta \nu$ range, whereas for several low $\Delta \nu$ disputed cases, the input population labels were either incorrect or ambiguous based on visual inspection. 
	
	We also made predictions on 7655 \textit{Kepler} red giants that do not have their asymptotic period spacing measured, from which 5379 have not been previously classified by any other means. We observed good agreement with the expected distribution of red giant populations in $\epsilon-\Delta \nu$ space for $\Delta \nu > 2.8\mu$Hz. Despite being currently limited to predicting within the asteroseismic parameter ranges of the training set, this new, simple,  and effective method of classifying oscillation spectra seems promising for further future classifications on large datasets in asteroseismology.
	
	\section*{Acknowledgements}
	Funding for this Discovery mission is provided by NASA's Science Mission Directorate. We thank the entire \textit{Kepler} team without whom this investigation would not be possible. D.S. is the recipient of an Australian Research Council Future Fellowship (project number FT1400147). We would also like to thank Timothy Bedding, Daniel Huber, and the group at The University of Sydney for fruitful discussions.
	
	%%%%%%%%%%%%%%%%%%%%%%%%%%%%%%%%%%%%%%%%%%%%%%%%%%
	
	
	
	\bibliographystyle{mnras}
	\bibliography{bibi} % if your bibtex file is called example.bib


	%%%%%%%%%%%%%%%%%%%%%%%%%%%%%%%%%%%%%%%%%%%%%%%%%%
	
	%%%%%%%%%%%%%%%%% APPENDICES %%%%%%%%%%%%%%%%%%%%%

	%\begin{figure}
	%	\centering
	%	\scalebox{0.4}{\input{/home/z3384751/Downloads/mnras/Diagram2.tex}}	
	%	\caption{Box}
	%\end{figure}
	
	%%%%%%%%%%%%%%%%%%%%%%%%%%%%%%%%%%%%%%%%%%%%%%%%%%

	% Don't change these lines
	\bsp	% typesetting comment
	\label{lastpage}
\end{document}